\documentclass[prb,twocolumn,superscriptaddress,showpacs,floatfix]{revtex4-1}
\usepackage{graphicx,epsfig,psfrag}
\usepackage{epstopdf}
\usepackage{amsfonts}
\usepackage{appendix}
\usepackage{graphics,dcolumn,bm,epic,eepic,float}
\usepackage{amssymb,amsmath,rotate,color}
\usepackage[colorlinks=true,linkcolor=blue,urlcolor=blue,citecolor=blue,anchorcolor=blue]{hyperref}
\setcitestyle{numbers,square}
\usepackage[normalem]{ulem}

\begin{document}
\newcommand{\be}{\begin{equation}}
\newcommand{\ee}{\end{equation}}
\newcommand{\la}{\langle}
\newcommand{\ra}{\rangle}
\newcommand{\bra}[1]{\ensuremath{\left\langle#1\right|}}
\newcommand{\ket}[1]{\ensuremath{\left|#1\right\rangle}}
\newcommand{\bracket}[2]{\ensuremath{\left\langle#1 \vphantom{#2}\right| \left. #2 \vphantom{#1}\right\rangle}}

\title{Identification of topological superconductivity in magnetic impurity systems using bulk spin-polarization}
\date{\today}

\author{Mahdi Mashkoori}
\affiliation{Department of Physics and Astronomy, Uppsala University, Box 516, SE-751 20 Uppsala, Sweden}
\affiliation{Department of Physics, K.N. Toosi University of Technology, P. O. Box 15875-4416, Tehran, Iran}

\author{Saurabh Pradhan}
\affiliation{Department of Physics and Astronomy, Uppsala University, Box 516, SE-751 20 Uppsala, Sweden}
\affiliation{Institut f\"{u}r Experimentalphysik, Freie Universit\"{a}t Berlin, Arnimallee 14, 14195 Berlin, Germany}

\author{Kristofer Bj\"{o}rnson}
\affiliation{Department of Physics and Astronomy, Uppsala University, Box 516, SE-751 20 Uppsala, Sweden}
\affiliation{Niels Bohr Institute, University of Copenhagen, DK-2100 Copenhagen, Denmark}

\author{Jonas Fransson}
\affiliation{Department of Physics and Astronomy, Uppsala University, Box 516, SE-751 20 Uppsala, Sweden}

\author{Annica M. Black-Schaffer}
\affiliation{Department of Physics and Astronomy, Uppsala University, Box 516, SE-751 20 Uppsala, Sweden}

\begin{abstract}
Magnetic impurities on the surface of Rashba spin-orbit coupled, but otherwise conventional, superconductors provide a promising way to engineer topological superconductors with Majorana bound states as boundary modes. In this work we show that the spin-polarization in the interior of both one-dimensional impurity chains and two-dimensional islands in these systems can be used to determine the superconducting topological phase, as it changes sign exactly at the topological phase transition. Thus, spin polarization offers an alternative method to detect the topological phase in magnetic impurity chains and islands deposited on conventional superconductors, beyond the zero energy Majorana bound states.
\end{abstract}

\maketitle
%
%
\section{Introduction}
Topological states of matter have been at the center of attention in condensed matter physics for the past decade \cite{Kitaev01,Wilczek09,BeenakkerRMP,Leijnse12,Aguado17}. The notion of topology as a classifier uses non-local properties, such as the Berry phase, and is thus fundamentally different from the traditional Landau-Ginzburg paradigm for phase transitions using local order parameters \cite{BernevigBook,WenBook}. In terms of realizing topological superconductivity, different platforms have already been proposed, such as spin-orbit coupled nanowires in proximity to superconductors or nanostructures created by depositing magnetic atoms on the surface of superconductors \cite{Anindya2012,Mourik2012, Churchil2013, Yazdani14,  Pawlak2015, Ruby17,Menard17, Andolina17,Wiesendanger18,Palacio18,Steinbrecher2018,Choi19}. 

According to the bulk-boundary correspondence \cite{Sato17}, zero-energy Majorana bound states (MBSs) appear at the end-points of many one-dimensional (1D) topological superconductors. The appeal of MBSs is particularly strong considering that they might be suitable for topological quantum computation \cite{Bravyi2002,Yu2003}. Several experiments have already reported zero-energy peaks in both impurity chains and nanowires \cite{Mourik2012,Yazdani14,Pawlak2015,Wiesendanger18}, suggestive of non-trivial topology and MBSs. For impurity chains, finite spin-polarization of the MBSs has further been used to differentiate MBSs from trivial states  \cite{Jeon17,Li18}. 

In this work we consider both 1D impurity chains and 2D impurity islands deposited on spin-orbit coupled but otherwise conventional $s$-wave superconductors and show that the spin-polarization of the low-energy states, measured in the interior, or bulk, and along the direction of the magnetic impurity moments, can also be used to determine the topological state of the system. These low-energy states are formed from hybridizing Yu-Shiba-Russinov (YSR) states \cite{Yu1965,Shiba1968,Rusinov1969}, which arise within the energy gap for magnetic impurities in conventional superconductors. An individual YSR state has a spin-polarization whose direction is set by its energy being positive or negative. We show that for both ferromagnetic chains and islands deposited on conventional superconductors with large spin-orbit coupling (SOC), this spin-polarization remains and encodes the topological phase transition (TPT) as an interchange of the spin-polarization between negative and positive low-energy states. We also show that the same interchange takes place for impurity chains with helical or other more complicated spin structures if using the locally defined magnetic moment direction as the basis for the spin-polarization. Our result might be extendable to nanowire systems as well.
Thus our findings establish that the spin-polarized local density of states (SP-LDOS), measurable using spin-polarized scanning tunneling spectroscopy (STS) \cite{Cornils17,Guigou16,Szumniak17,Chevallier18}, provides a powerful tool to verify non-trivial topology in the superconducting phase for magnetic impurity systems on conventional superconductors, entirely independent of the existence of MBSs.

%
\section{Model}
\begin{figure}[t]
\centering
\includegraphics[width=0.9\columnwidth]{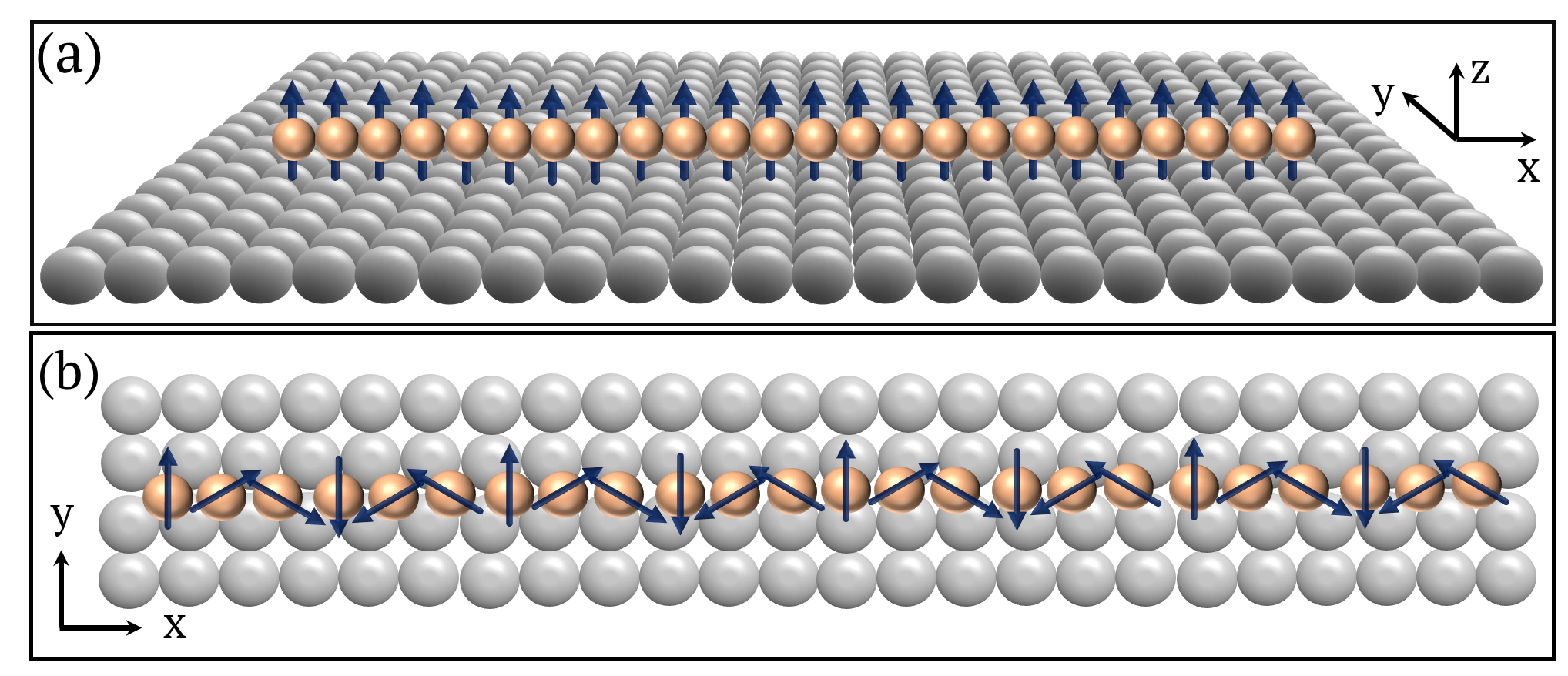}
\caption{Schematic illustrations of impurity chains on a superconductor: (a) ferromagnetic impurity chain with $z$-axis magnetic moments (FMC) and (b) top view of spin-helical impurity chain (SHC).}
\label{schematics.fig}
\end{figure}

To avoid unnecessary complexities, we keep our model simple yet capturing all important features. For the substrate we consider a square lattice, lattice constant $a=1$, with Rashba SOC and conventional $s$-wave superconductivity. The full mean-field Hamiltonian reads $ {\cal H} = {\cal H}_{\rm sub} + {\cal H}_{\rm imp}$ where,
\begin{subequations}
\label{tight-binding.eq}
\begin{align}
{\cal H}_{\rm sub} &=
-\frac{1}{2}\sum_{ {\bf ij} \alpha}
t_{{\bf ij}} c_{{\bf i}\alpha}^\dag
c_{{\bf j}\alpha}
+ \sum_{ {\bf i} }
({\Delta }c_{{\bf i}\uparrow}^\dag
c_{{\bf i} {\downarrow}}^\dag + \textrm{H.c.})
\nonumber \\ 
& -\lambda_R \sum_{{\bf i} ,r=\pm} r c_{{\bf i},\uparrow}^{\dag} 
(c_{{\bf i}-r\hat{\bf x},\downarrow}-
ic_{{\bf i}-r\hat{\bf y},\downarrow}) + \textrm{H.c.},
\\
 {\cal H}_{\rm imp} & = 
\sum\limits_{\textbf{i}\alpha \beta}  Jc_{\textbf{i} \alpha}^{\dag} \vec{{S}}_{\textbf{i}} \cdot  \vec{\sigma}_{\alpha \beta}  c_{\textbf{i} \beta}.
\end{align}
\end{subequations}
Here $c_{\bf i \alpha}(c^\dagger_{\bf i \alpha})$ is the creation (annihilation) operator at site $\textbf{i}=(i_x,i_y)$ with spin $\alpha \in \{\uparrow,\downarrow\} $ and $\vec{\sigma} = (\sigma_x,\sigma_y,\sigma_z)$ with $\sigma_{i}$ the Pauli spin matrices. 
The chemical potential is $t_{{\bf i}{\bf i}}=2\mu$ \cite{Two} and we restrict the range of hopping to nearest neighbors: $t_{\bf{i}\neq {\bf j}} = t =1$. Rashba SOC is present due to inversion symmetry breaking at the surface and set by $\lambda_R$, while superconductor order is given by $\Delta$. 
Furthermore, magnetic impurities behaving as classical moments \cite{Nadj-Perge2013,Pientka2013}, such that $J\vec{S}$ mimics a local Zeeman field $V_{Z}\hat{n}$ on each impurity site \cite{Shiba1968}. 
Inspired by different experimental setups, we study both ferromagnetic chains with all moments perpendicular to the substrate ($\hat{n} = \hat{z}$) and spin-helical chains where the moments lie in $x$-$y$ plane ($\hat{n} \perp \hat{z}$), see Fig~\ref{schematics.fig}. We also consider ferromagnetic 2D impurity islands. 
Without loss of generality we assume that the impurity chains are oriented along the $x$-axis while the impurity island forms a circle, with both systems embedded in the middle of a large square lattice.  
We are here primarily concerned with the SP-LDOS: $\rho_{i}({\bf{i}},E) = \la \psi^\dagger_{\bf{i}} {{\sigma}_{i}} \psi_{\bf{i}} \ra$, with $\psi_{\bf{i}}^\dagger = (c_{\bf{i}\uparrow}^\dagger, c^\dagger_{\bf{i}\downarrow})$ and measurable using STS with spin-polarized tips, calibrated along a specific spin direction \cite{Cornils17}.
We calculate the SP-LDOS within a Bogoliubov-de Gennes (BdG) formulation of Eq.~\eqref{tight-binding.eq} by using a Chebyshev expansion of the Green's function \cite{Weibe06,Covaci2010,Bjornson2016,Mashkoori2017,Bjornson2019}.

%
%

\section{Results}
\subsection{1D ferromagnetic system} 
To understand the behavior of the spin-polarization, we start by studying a pure 1D ferromagnetic system, which we obtain by shrinking the superconducting substrate along the $y$-axis to the width of one single unit cell. We also apply periodic boundary conditions in the $x$-direction and Fourier transform to arrive at the 1D BdG Hamiltonian ${\cal H} = \sum_{k} \Psi_{k}^\dag {\cal H}_{\rm 1D}(k) \Psi_{k}$ with 
\begin{equation}
{\cal H}_{\rm 1D}(k) =  {\tau_{z}}\left( {{\xi _k}{\sigma_{0}} + L_R \sigma_{y} } \right) + {V_Z}{\tau_{0}}{\sigma_{z}} + \Delta {\tau_{x}}{\sigma_{0}},
\label{BdG.eq}
\end{equation}
and Nambu spinor $\Psi_{k}^T = (c_{k \uparrow },c_{k \downarrow },c_{-k \downarrow }^\dag , - c_{-k \uparrow }^\dag)$.
Here $\sigma_i$ $(\tau_i)$ are Pauli matrices in spin (particle-hole) space, the SOC is $L_R=2\lambda_R\sin{k_x}$, normal band dispersion $\xi_k=-2t\cos{k_x}-\mu$, and $V_Z$ the impurity-induced uniform Zeeman field in the $+\hat{z}$-direction. Diagonalizing this Hamiltonian, we find four bands: 
$E=\pm\sqrt{ \xi _k^2 + {L_R^2} + V_Z^2 + {\Delta ^2} \pm 2\sqrt {\xi _k^2\left( {{L_R^2} + V_Z^2} \right) + V_Z^2{\Delta ^2}}}$.
Since $\Delta$ is $k$-independent, it is easy to show that $L_R$ has to vanish at the closing points of the energy gap \cite{Sato2006,Sato2010}. Thus, gap closings occur at the high symmetry points $k_x \in \left \{ 0, \pm \pi \right \}$ and for $V_Z^{c_{\pm}} = \sqrt{(|\mu| \pm 2)^2+\Delta^2}$. These gap closings are TPTs with a topological phase appearing between $V_Z^{c_-}$ and $V_Z^{c_+}$ \cite{Sato2010}.
%
\begin{figure}[t]
\centering
\includegraphics[width=\columnwidth]{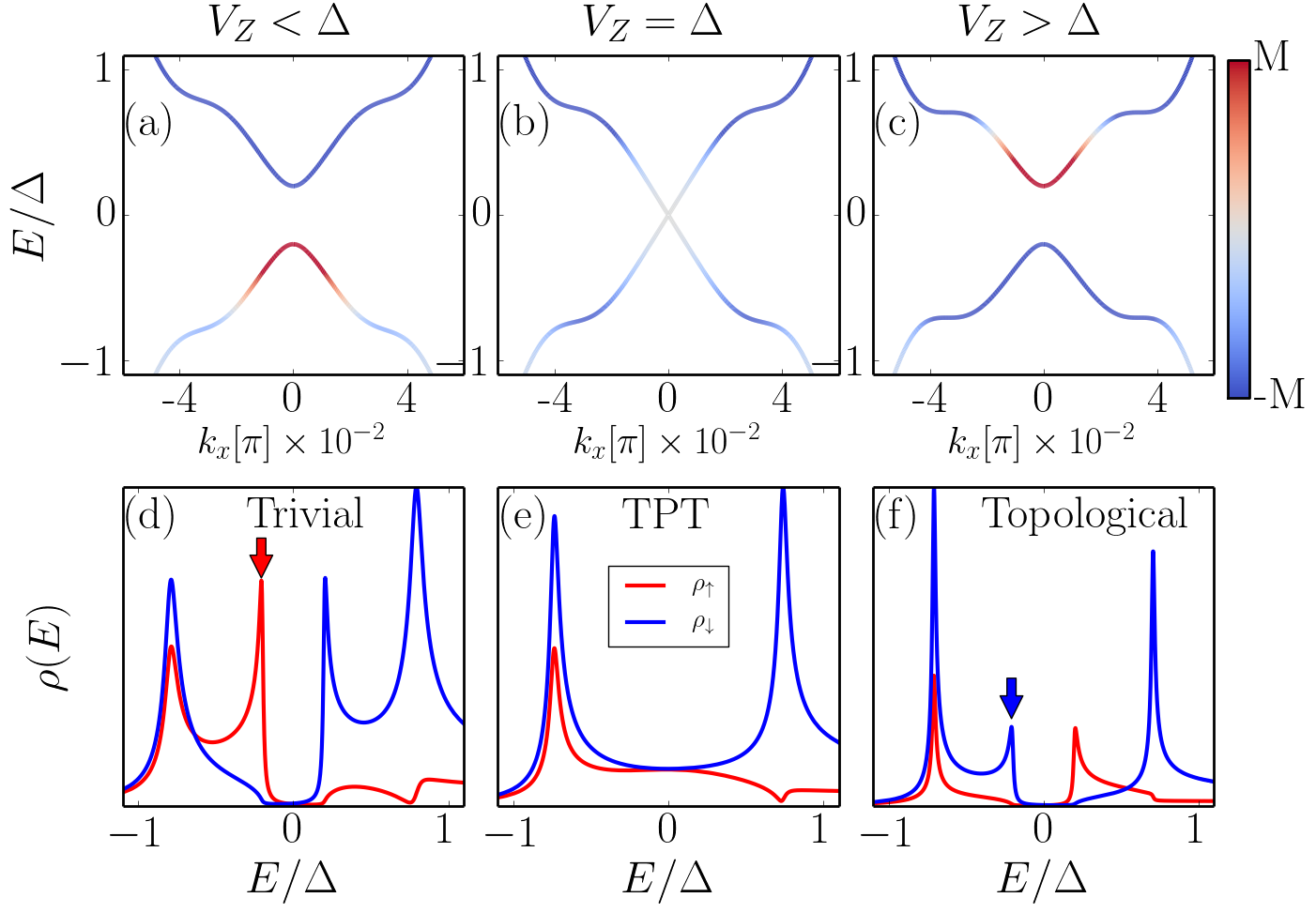}
\caption{Low-energy band structure for 1D ferromagnetic system (a-c) and corresponding SP-DOS (d-f) in trivial (left), at TPT (middle), and topological (right) phase with $\pm\hat{z}$ polarization (red/blue). Band color represents spin-polarization along the $z$-axis, with each plot individually renormalized, while arrows indicate the spin-polarization switching at the TPT for the negative energy band. Here $\Delta = 0.01$, $\lambda = 0.1$ and $\mu = -2$. }
\label{Fig2.fig}
\end{figure}

In Fig.~\ref{Fig2.fig}(a-c), we plot the two lowest energy bands, also referred to as YSR bands due to their impurity origin, as a function of $V_Z$ across the lowest TPT. We tune $V_Z$ as the exact value of $V_Z$ is usually unknown and is also experimentally tunable \cite{Hatter2015,Farinacci2018,Wiesendanger18}. We first choose a chemical potential at the bottom of the normal state band, $\mu = -2$. Then, the superconducting gap opens in the vicinity of $k_x = 0$, where also the first TPT occurs at $V_Z^{c_-}=\Delta$. 
In the trivial phase, $V_Z<\Delta$, we find a spin-polarization close to $k_x=0$, with the negative and positive energy YSR bands being spin-polarized (same as impurity moment) along the $\hat{z}$ and $-\hat{z}$ directions, respectively. At the TPT, the energy gap closes and the spin-polarization vanishes at the lowest energies, but in the topological phase, $V_Z>\Delta$, the $z$-axis spin-polarization of the positive and negative bands is interchanged. As a consequence, the SP-DOS along the $z$-axis, plotted in Fig.~\ref{Fig2.fig}(d-f), has a positive peak (spin-up, red) for negative energies in the trivial phase but a negative peak (spin-down, blue) in the topological phase, clearly showing how the SP-DOS is interchanged at the TPT and thus offering an easily measurable signature of the TPT. 

A physical explanation of the spin interchange between the lowest order energy bands at the TPT is conceived by recalling that each magnetic impurity in a conventional superconductor induces a pair of fully spin-polarized YSR states \cite{Cornils17}. For weak impurity moments, the spin-polarization of a single YSR state with negative (positive) energy is aligned (anti-aligned) with the moment of the impurity. Increasing the moment strength, a quantum phase transition takes place and the spin-polarization of the negative and positive energy states is interchanged \cite{BalatskyRMP}. 
When arranging magnetic impurities into a chain, the YSR states starts to overlap and instead form two fully spin-polarized YSR bands. In the absence of SOC, the same spin-interchange effect as for the single impurity YSR states appears also for the YSR bands. However, finite SOC leads to an admixture of spin-up and spin-down states. Still, spin is good quantum number at the high symmetry points $k_x \in \{0, \pm \pi \}$, since the SOC contribution vanishes at these points. Thus, close to $k_x = 0$, and thus at the TPT, we expect a spin-interchange for a chain, as also seen in Fig.~\ref{Fig2.fig}. 

\begin{figure}[t]
\centering
\includegraphics[width=\columnwidth]{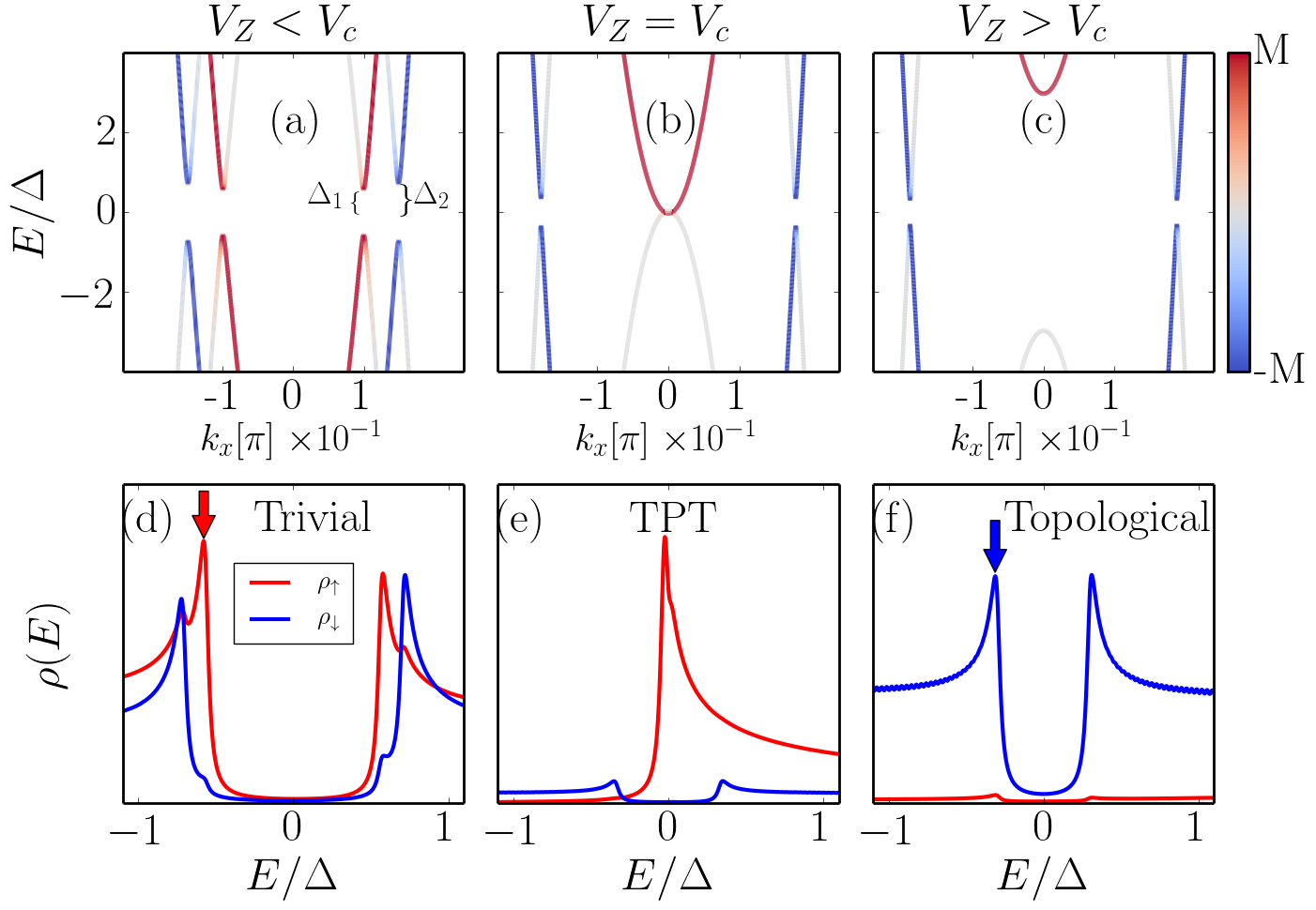}
\caption{Same as Fig.~\ref{Fig2.fig}, except for finite doping, $\mu = -1.85$.}
\label{Fig3.fig}
\end{figure}

The chemical potential is, however, often not at the bottom of the band and we depict a more general situation in Fig.~\ref{Fig3.fig} for finite doping. Here, inner and outer band gaps are typically found, labeled by $\Delta_1$ and $\Delta_2$ in Fig.~\ref{Fig3.fig}(a), which are attributed to two helical bands. 
Starting from small Zeeman fields, the inner gap $\Delta_1$ shrinks and eventually closes at a TPT at $k_x=0$ for increasing $V_Z$, see Fig.~\ref{Fig3.fig}(b), while $\Delta_2$ remains essentially unaffected \cite{Three}. 
In the trivial phase, we find dominant spin-up polarization for the lowest-energy states, which at the TPT even generates a single peak. The latter is due to the negative energy YSR band having a completely hole-like spin-down component around $k_x=0$ at the TPT, thus giving no contribution to the DOS.
To quantify this behavior and compare it to Fig.~\ref{Fig2.fig}, we study the Hamiltonian Eq.~\eqref{BdG.eq} at $k_x=0$ for a general $\mu$. We find that the spin-polarization of negative and positive energy YSR bands at $k_x=0$ follows from the ratio of $\eta=|{2+\mu}|/{\Delta}$. Whenever $\eta < 1$, as in Fig.~\ref{Fig2.fig}, we find an electron-like behavior for both bands around the TPT. However, for $\eta > 1$, as in Fig.~\ref{Fig3.fig}, the spin-down (spin-up) states become completely hole-like (electron-like), see \ref{App1} for  details.

Tuning $V_Z$ further into the topological phase, $\Delta_1$ opens again and rapidly becomes larger than $\Delta_2$. Therefore, in the topological phase, the sharp spin-up peak moves to higher positive energies, while only spin-down polarization remains at lower energies. Most notably, at negative energies the SP-LDOS is always spin-down polarized beyond the TPT, since the states associated with $\Delta_1$ are always hole-like for negative energies. Thus, the TPT is inherently connected with an interchange of the bulk spin-polarization for the lowest negative energy bands, as schematically indicated with colored arrows in Fig.~\ref{Fig3.fig}(d-f) and in full agreement with the earlier results in Fig.~\ref{Fig2.fig}.
Further increasing $V_Z$, the $\Delta_2$ gap eventually closes at $k_x=\pm \pi$ in a second TPT, also with a spin-interchange of the YSR bands, see \ref{App2} for detials.

\begin{figure}[t]
\centering
\includegraphics[width=\columnwidth]{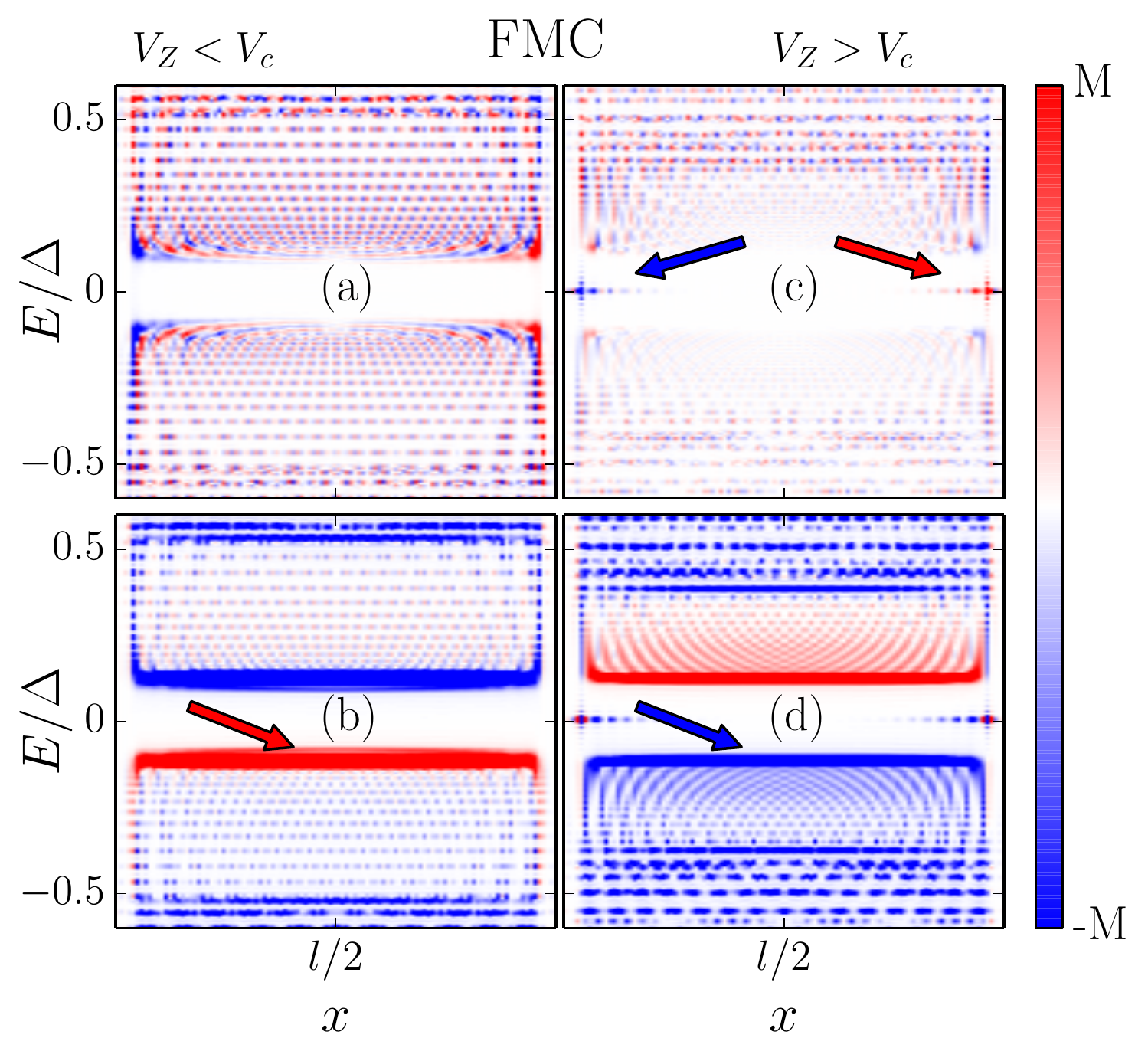}
\caption{SP-LDOS along a FMC (a-d) with length $l = 101$. Upper row shows $x$-axis spin-polarization, lower row $z$-axis spin-polarization. Arrows indicate the spin-polarization of the lowest energy states in the bulk, signaling the TPT, and $x$-axis spin-polarization of the MBSs.
Here $\mu = -2$, $\Delta = 0.4$, $\lambda_R=0.4$, while $V_Z=1.9 (2.4)$ for (non)-trivial phases.}
\label{FMChain.fig}
\end{figure}
 
\subsection{Ferromagnetic impurity chain }
\label{subsec:FMC}
Having understood the pure 1D limit, we next perform numerical calculations for one of the systems we set out to study: a finite ferromagnetic impurity chain (FMC) embedded in a 2D superconducting substrate, see Fig.~\ref{schematics.fig}(a). The superconducting substrate consists of $L_\parallel = 201$ lattice sites in the direction of the $l = 101$ long chain, and $L_\perp=21$ sites perpendicular to the chain. We here set $\mu = -2$, which puts the 2D system well within a finite doping regime. 
For visualization purposes, we set $\Delta=0.2$ or $0.4$ in all calculations, however, the same conclusions hold for smaller values.
In Fig.~\ref{FMChain.fig} we show both the $x$- and $z$-axis spin-polarization in the trivial and topological phases. In the topological phase we find MBSs at the end-points of the chain with a significant $x$-axis spin-polarization, in agreement with earlier results \cite{Sticlet2012,Bjornson2015,Jeon17}.

Beyond the MBSs spin-polarization, we also find strong $z$-axis spin-polarization of the in-gap YSR states in the central regions of the chain. 
Focusing on the lowest energy YSR states, we see in Fig.~\ref{FMChain.fig}(b) that in the trivial phase the negative energy subgap states are dominantly spin-polarized along the $\hat{z}$-direction, i.e.~parallel to the impurity spin (red), while positive energy states are mostly aligned along $-\hat{z}$-direction (blue). Remarkably, in the topological phase this spin-polarization is reversed, see Fig.~\ref{FMChain.fig}(d). 
We here point out that in the topological phase the spin-polarization of the positive energy YSR states ultimately depends on the Zeeman field as these states can again switch spin-polarization with increasing $V_Z$, as explained in \ref{App2}. However, the negative energy YSR states are always anti-parallel to the impurity moment in the topological phase, as also indicated by the colored arrows in Fig.~\ref{FMChain.fig}(b,d). Thus, the spin-polarization in the chain interior of the lowest negative energy YSR states becomes a probe of the bulk topology.
\begin{figure}[t]
\centering
\includegraphics[width=\columnwidth]{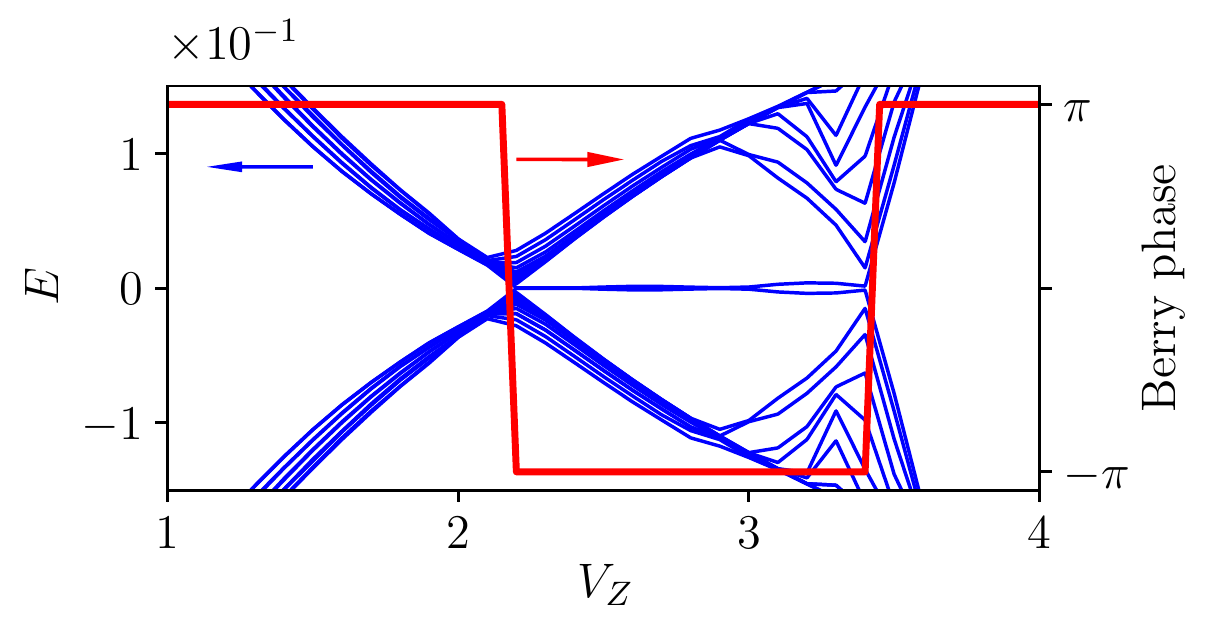}
\caption{Dispersion of a few of the lowest YSR subgap states for a FMC as a function of $V_Z$ (left axis) and the corresponding Berry phase evaluated for a nano-ribbon of width $l_y = 31$ lattice point (right axis). Other parameters are the same as in Fig.~\ref{FMChain.fig}.}
\label{FigNew.fig}
\end{figure}

As another indicator of the close relation between the bulk topology and spin-polarization of lowest negative energy YSR states, we plot a few of the lowest YSR subgap states in Fig.~\ref{FigNew.fig} and the corresponding Berry phase of the occupied bands as a topological index. To do so, we Fourier transform the Hamiltonian along the chain and perform Wilson-loop characterization  of the occupied bands \cite{Alexandradinata14,Bouhon19}. As illustrated in Fig.~\ref{FigNew.fig}, at exactly the same lower critical coupling $V_Z= 2.1$, the Berry phase sharply drops from $+\pi$ to $-\pi$, signaling the TPT. Then, the Berry phase jumps from $-\pi$ to $+\pi$ at the second TPT at $V_Z = 3.4$ where the FM impurity chain becomes trivial again. This further establishes the topological non-trivial regime for intermediate magnetic couplings.

\subsection{Spin-helical impurity chain}
\begin{figure}[t]
\centering
\includegraphics[width=\columnwidth]{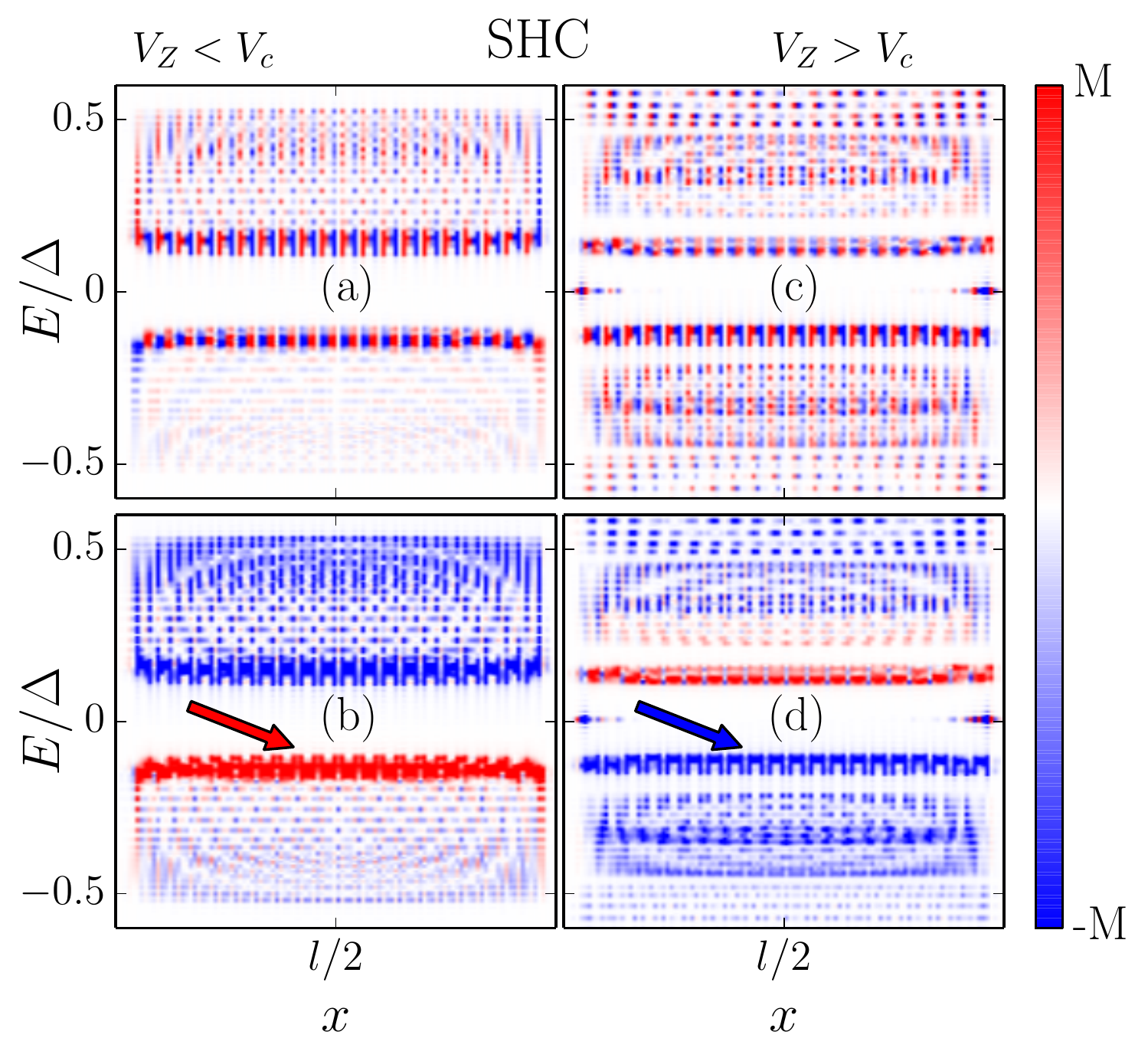}
\caption{SP-LDOS along a SHC (a-d) lengths $l = 101$. Upper row shows $x$-axis spin-polarization, lower row spin-projected (on spin-helix) LDOS for the SHC. Arrows indicate the spin-polarization of the lowest energy states in the bulk, signaling the TPT.
Parameters are the same as in Fig.~\ref{FMChain.fig}, except we set $\lambda_R = 0.2$ for SHC.}
\label{SHChain.fig}
\end{figure}

Next we discuss a spin-helical impurity chain (SHC), also likely experimentally realized \cite{Pawlak2015,Schecter16,Christensen16,Steinbrecher2018}. As illustrated in Fig.~\ref{schematics.fig}(b), for an impurity located at $x_{\bf{i}}$ the local moment is in-plane and given by $\vec S\left( {\bf{i}} \right) = \left( S\cos \left( {{k_h}{x_{\bf{i}}}} \right),  S\sin \left( {{k_h}{x_{\bf{i}}}} \right), 0 \right)$, where ${k_h} = 2\pi /\ell$ with $\ell$ being the pitch of the spin-helix \cite{One}. In Fig.~\ref{SHChain.fig}(a,c) we plot the $x$-axis SP-LDOS, which demonstrate how the MBSs appear in the topological phase at the end-points of the chain, but notably their spin-polarization is no longer constant. 

With the helical spin structure, the spin-texture of the low-lying YSR states in the chain interior are alternating between up and down directions for $x,y$-axes SP-LDOS, following the pitch of the spin-helix of the implanted magnetic impurities. 
However, motivated by the fact that a SHC is topologically equivalent to a FMC plus an additional SOC \cite{Klinovaja2013}, we find a way to map the SP-LDOS and still identify the TPT: We evaluate the SP-LDOS along the SHC where at each lattice point $\bf{i}$ the spin-polarization is projected on $\vec{S}(\bf{i})$: 
$
{\rho _n}\left( {\bf{i},E} \right)~=~\cos({{k_h}{x_{\bf{i}}}}){\rho _x}\left( {\bf{i},E} \right)~+~\sin({{k_h}{x_{\bf{i}}}}){\rho _y}\left( {\bf{i},E} \right).
$
As illustrated in Fig.~\ref{SHChain.fig}(b,d) this spin-projection is successful in providing a clear spin-polarization signature of the TPT. Concentrating on the low-energy YSR states at negative energies, this spin-projected SP-LDOS changes from being dominantly spin-up (red) in the trivial phase to spin-down (blue) in the topological phase. Thus spin-projected SP-LDOS for the SHC can be used in exactly the same way as the out-of-plane SP-LDOS for the FMC in predicting the topological phase only based on bulk signatures, see colored arrows in Fig.~ref{SHChain.fig}(b,d), and notably fully independent from the existence of the MBSs. Clearly, the same spin-projection procedure is capable of handling arbitrarily complicated spin structures in the chain: the relevant spin-polarization direction for predicting the TPT is always defined by the orientation of each impurity moment in the normal (non-superconducting) phase and thus experimentally accessible.

\subsection{Dilute FMCs}
In order to provide results for varying inter-impurity distances we also perform a T-matrix analysis based on an equivalent continuum model for FMCs where we can easily vary the inter-impurity distances. These results goes beyond the nearest neighbor distance used in the lattice calculations above, and provide results for more dilute FMCs. 
The T-matrix formalism is efficient when the number of impurities is small, or at least discrete, and embedded in a continuum.  Generally, we write the Green's function of the system  $H = H_0 + V$ as
 \begin{align}
 & G  = (\omega -H)^{-1}
   = G_0 + G_0 T G_0, \nonumber\\
&G(r_i,r_j,\omega)  = G_0(r_i-r_j,\omega)
\\&
	  + G_0(r_i-r_k,\omega) T(r_k,r_l,\omega) G_0(r_l-r_j,\omega) .
\nonumber
 \end{align}
Here, $H_0$ and $G_0$ are the Hamiltonian and Green's function of the system without impurities, respectively, and $T = (V^{-1}-G_0)^{-1}$ is the $T$-matrix, which includes all the effects of the impurities encoded in $V$. In the above equations, all the elements are matrices for a multi-impurity system.   
In particular, we consider a superconducting substrate with SOC. The substrate Hamiltonian in the Nambu basis can be written as 
\begin{align}
\label{H0.eq}
H_0 = 
\begin{pmatrix}
{\xi_k}\sigma_0& \Delta\sigma_0 \\
 \Delta\sigma_0 &-\xi_k \sigma_0
\end{pmatrix}
+H_{\rm SOC}
,
\end{align}
where $\xi_k$ is the normal-state dispersion relation for free electrons, $\Delta$ the conventional superconducting order parameter, and the SOC is modelled by  $ H_{\rm SOC} = \lambda \tau_0(p_y\sigma_x  - p_x\sigma_y )$. Here $\sigma$ and $\tau$ are the Pauli matrices in the spin and Nambu basis, respectively, with the latter explicitly written in matrix form in the first term of Eq.~\eqref{H0.eq}.
Similar to the lattice calculations in Section \ref{subsec:FMC}, we replace the effect of the impurities by an effective Zeeman field $V_Z$ along the $z$-axis for all magnetic impurities, thus ignoring dynamical processes such as the Kondo effect. The Hamiltonian for the impurities can therefore be written as 
\begin{align}
 V = {V}_Z \tau_0\sigma_z \delta(r).
\end{align}
Having thus defined $H = H_0+V$, the dressed Green's function $G(r_i,r_j,\omega)$ is provided in terms of the bare Green's function $G_0(r_i-r_j,\omega)$ through the T-matrix, where the bare propagator $G_0$ are expressed in terms of Hankel functions \cite{Kaladzhyan_JPCM,Kaladzhyan_PRB}. 

\begin{figure}[t]
\centering
\includegraphics[width=0.5\textwidth]{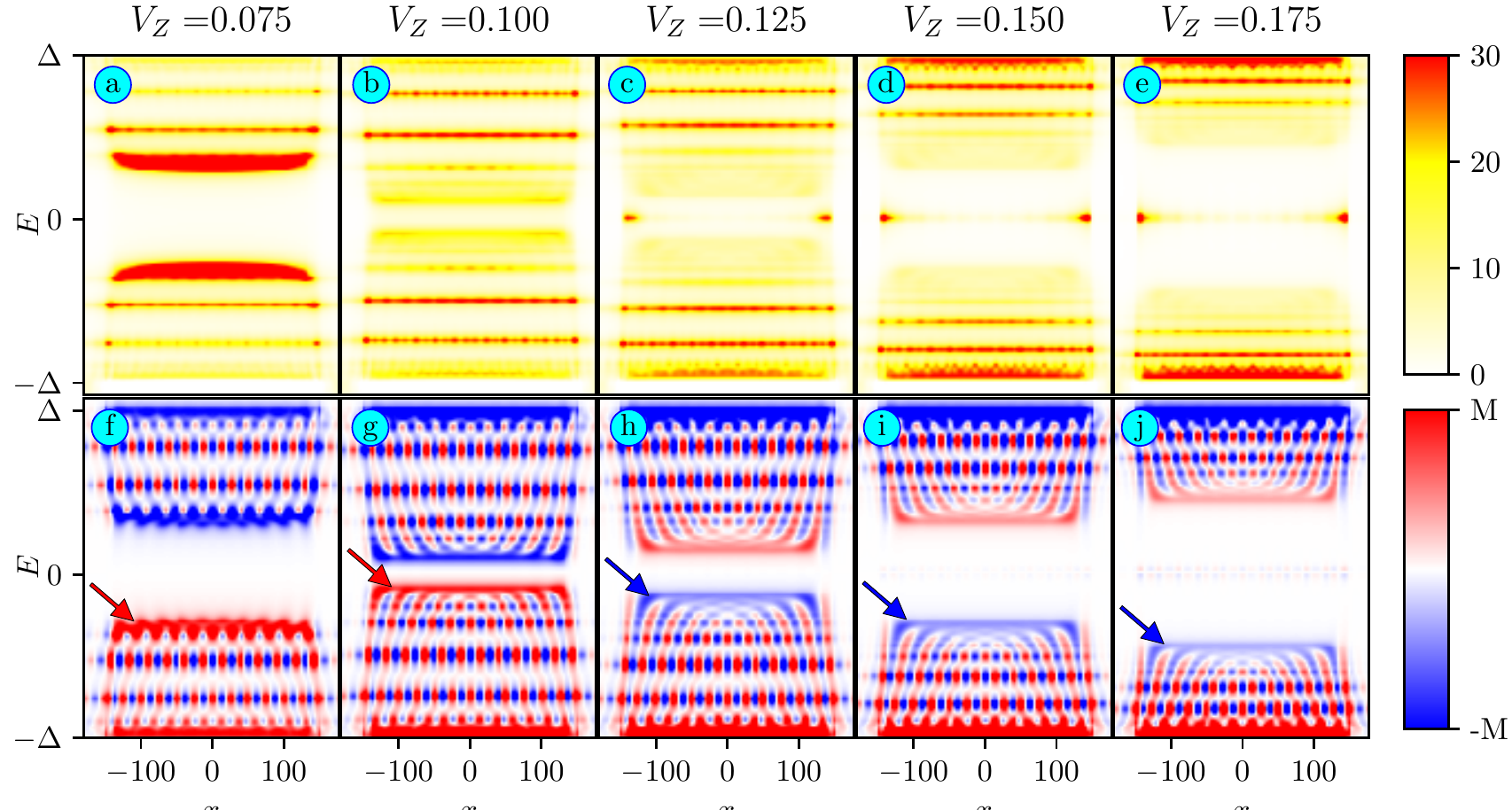}
\caption{LDOS (a-e) and $z$-axis spin-polarization (f-j) for a FMC within a continuum model. The chain consists of 301 impurities placed in-between -150 and 150 along the $x$-axis, giving an inter-impurity distance of $d \sim 5$. The Fermi velocity, to which we set the length scale, is $v_F=0.2$, superconducting order parameter $\Delta = 0.2$, and spin-orbit coupling $\lambda = 0.5$. Arrows indicate the spin-polarization of the lowest energy states in the bulk, signaling the TPT.}
\label{fig:tmatrix}
\end{figure}

For a chain with relatively small inter-impurity distances, i.e.~the dense limit with approximately unit size inter-impurity distances, we show the resulting LDOS and $z$-axis spin-polarization in Fig.~\ref{fig:tmatrix}. For small $V_Z$ the system remains in the trivial phase and the YSR states are gapped without any states emerging deep within the gap. By increasing $V_Z$, the YSR states approach the Fermi level and eventually cross each other and the system is driven through the TPT. In the topological phase MBSs emerge at the chain end-points, while the electronic structure remains gapped around the center of the chain. Following the $z$-axis spin-polarization we find in the trivial phase that the negative energy states (see red arrows) have a spin-up polarization, whereas positive energy states have spin-down polarization. 
By tuning the parameters such that the system goes through the TPT, the spin-polarization of the low-energy YSR states is thus interchanged. This result is exactly the same as for the lattice results in Fig.~\ref{FMChain.fig}, which models the ultimately dense limit (one magnetic impurity per lattice site).

\begin{figure}[t]
\centering
\includegraphics[width=0.5\textwidth]{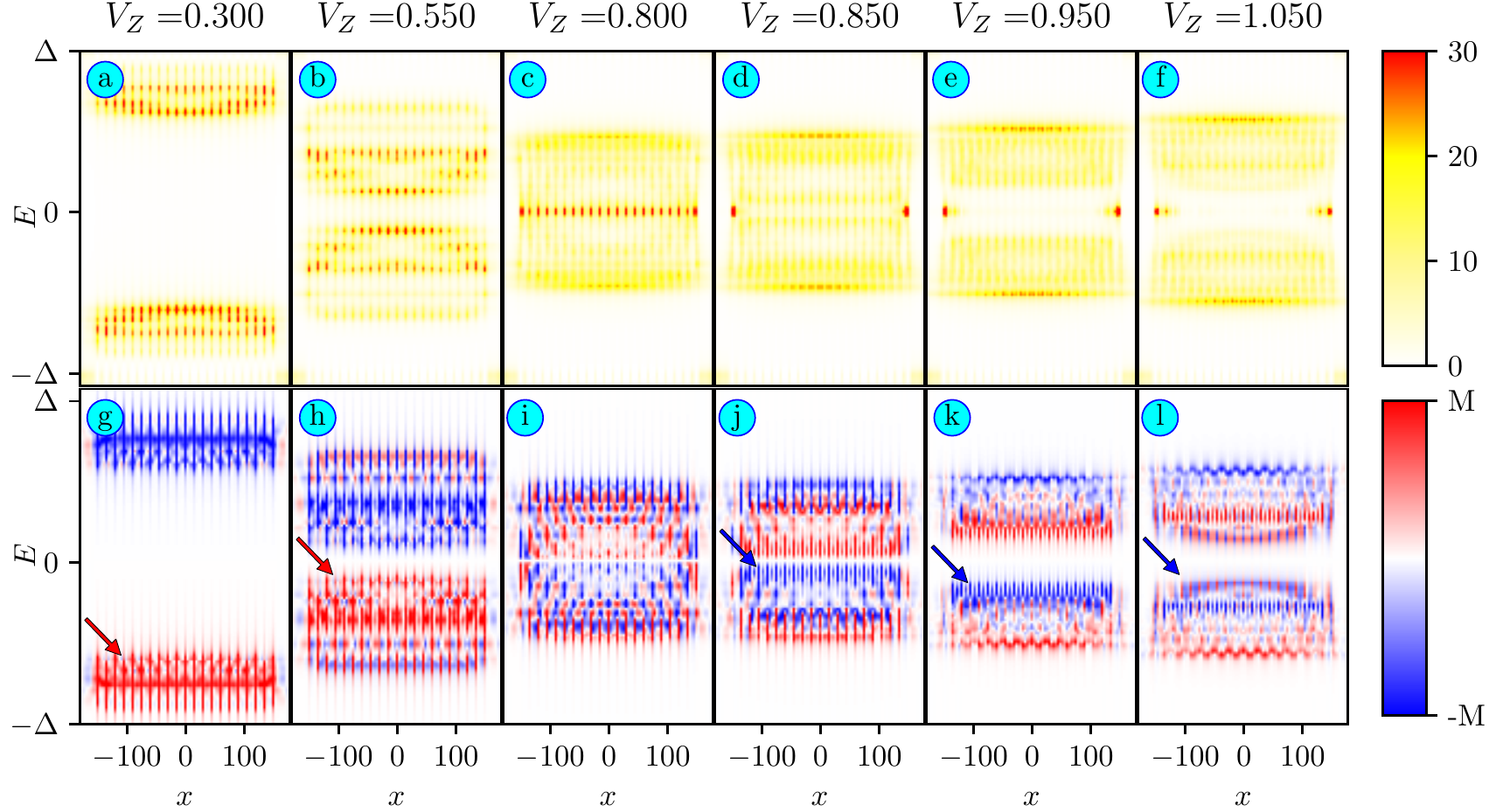}
\caption{Same as Fig.~\ref{fig:tmatrix} but for a chain consisting of 21 impurities, giving an inter-impurity distance of $d \sim 70$.}
\label{diluteFMC.fig}
\end{figure}

Performing the same calculations for more dilute FMCs, we ultimately arrive at the picture in Fig.~\ref{diluteFMC.fig}, where the inter-impurity distance is now approximately 14 times larger than in Fig.~\ref{fig:tmatrix}. We find that the spin-polarization still signals the TPT. However, the spin-polarization signal slightly fades away, a result due to the necessarily very weak hybridization of the YSR states in such dilute chains. The strength of the hybridization between individual YSR states is thus an important factor for providing clear bulk signatures of the TPT.
Overall, our T-matrix results illustrates how the spin-polarization pinpoints the TPT independent of the inter-impurity distance.

\subsection{Ferromagnetic impurity island}
\begin{figure}[hbt]
\centering
\includegraphics[width=\columnwidth]{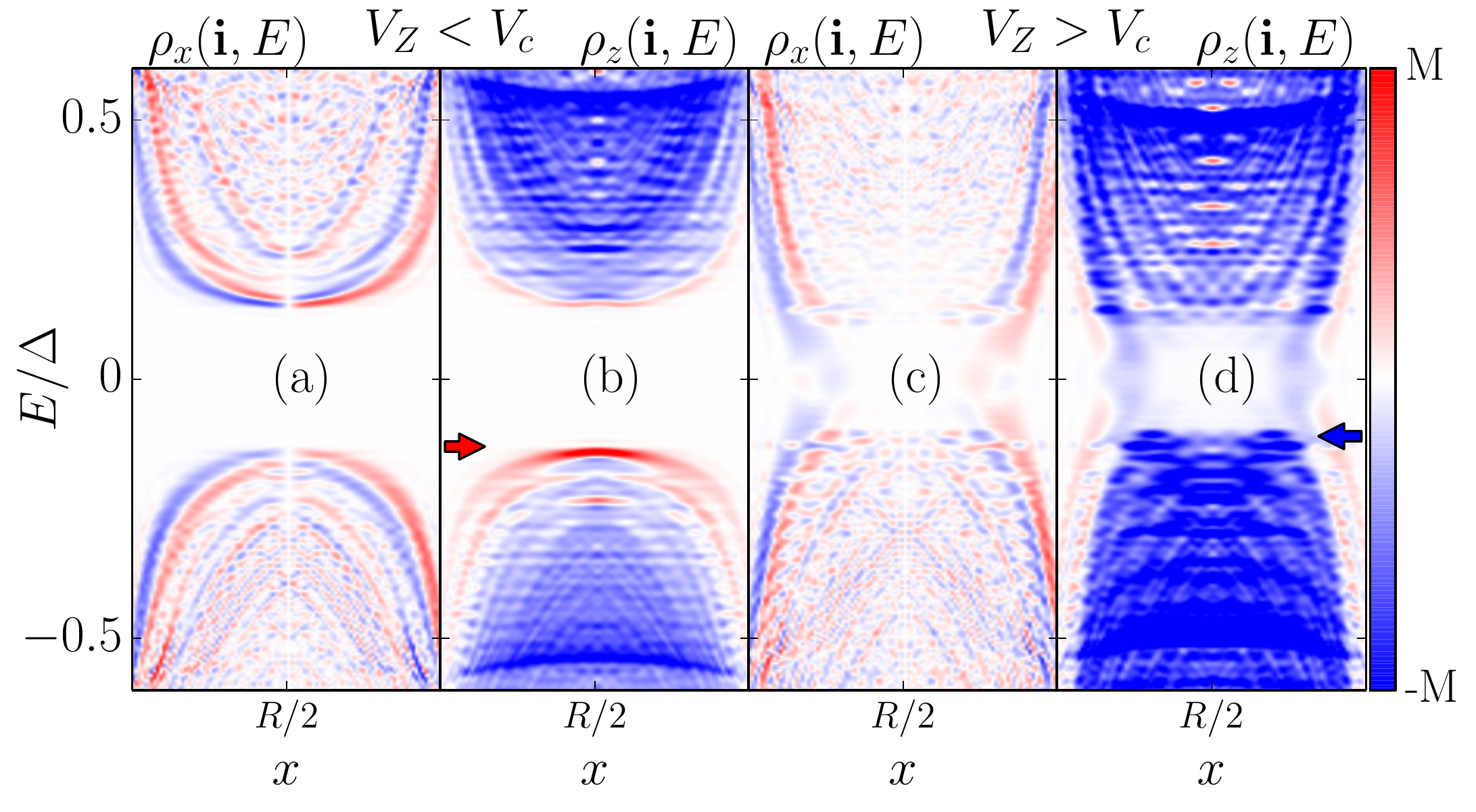}
\caption{SP-LDOS for across an island with radius $R = 50$ with $z$-axis ferromagnetic impurities for spin-polarizations along the $x$ (a,c) and $z$-axes (b,d). Here $V_Z = 2.1 (2.3)$ for (non-)trivial phases. Other parameters are the same as in Fig.~\ref{FMChain.fig}, but with $\mu = -2$ for a 2D island giving a high doping level.}
\label{Island.fig}
\end{figure}

Inspired by recent work on 2D impurity islands \cite{Menard17, Bjornson18,Palacio18}, we also study a ferromagnetic impurity island with all magnetic moments in the $\hat{z}$-direction on the surface of an $s$-wave superconductor with SOC. In Fig.~\ref{Island.fig} we show the SP-LDOS along a line through the impurity island for spin-polarizations along the $x$- (a,c) and $z$-axis (b,d) on both sides of the TPT. %
In the topological phase, chiral edge states appear at the island's boundary, with distinctive $x$-axis spin-polarization, in agreement with earlier results \cite{Bjornson18}. But most importantly, an interchange of the $z$-axis spin-polarization of the lowest energy YSR states at negative energies is present across the TPT in the middle of the island: in the trivial (topological) phase these YSR states are (anti-)aligned with the moment of the magnetic impurities, exactly the same as for 1D chains. This result is not limited to the particular parameter choices of Fig.~\ref{Island.fig}. In fact, assuming less doping in the normal state results in an even more pronounced and clear-cut spin-interchange signature for the TPT. Thus, measurements of the SP-LDOS along the magnetic impurity direction allows for determination of the topological superconducting phase for both 1D impurity chains and 2D islands. 

%
%

\section{Concluding Remarks}
In this work, we perform analytical and numerical calculations for 1D and 2D magnetic impurity structures adsorbed on the surface of a conventional superconductor with Rashba SOC. 
We find that for all dense impurity chains and islands the spin-polarization of the low-energy YSR states undergoes a spin-interchange across the topological phase transition. For dilute FMCs, the spin-polarization continues to signal the TPT, although the weak hybridization causes the signal to slightly fade away with increasing inter-impurity distance. 
Remarkably, recent SP-STS measurements in the middle of ferromagnetic Fe impurity chains on a conventional Pb superconductor, with putative MBS at the chain end points, showed dominance of spin-down (-up) LDOS at the negative (positive) low-energy states \cite{Jeon17}, in agreement with our results in the topological phase. Similar measurements for Co impurity chains, where no MBSs were found, showed the opposite spin-polarization \cite{Ruby17}, also in agreement with our identification of the trivial phase.
To conclude, we show how current spin-polarization measurements \cite{Jeon17, Ruby17} can be used as an additional tool, beyond the existence of any Majorana bound states, to determine the topological phase for magnetic impurity chains and islands absorbed on surfaces of conventional superconductors with Rashba spin-orbit coupling.
%
%
\acknowledgments
We acknowledge financial support from the Swedish Research Council (Vetenskapsr\aa det, Grant No.~621-2014-3721), the Swedish Foundation for Strategic Research (SSF), the Wallenberg Academy Fellows program through the Knut and Alice Wallenberg Foundation, and Stiftelsen Olle Engkvist Byggm\"astare.
The computations were performed on resources provided by the Swedish National Infrastructure for Computing (SNIC) at HPC2N and UPPMAX.

%
%
\appendix
\section{1D ferromagnetic system}
\label{App1}
\begin{figure}[h]
\center
\includegraphics[width=0.5\textwidth]{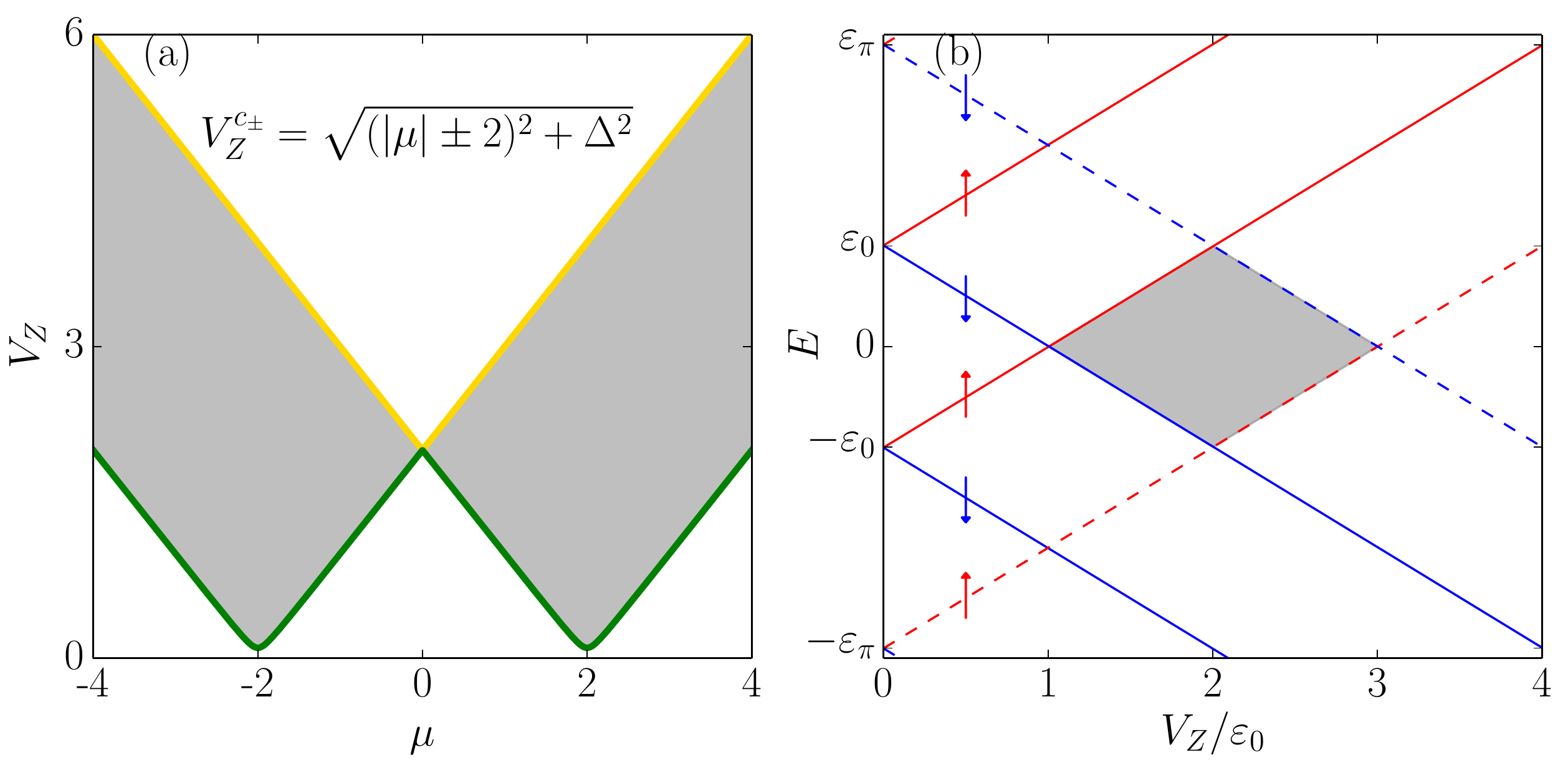}
\caption{Phase diagram of 1D ferromagnetic system (a) and energy of the YSR bands at high symmetry points of $k_x=0$ (solid) and $k_x=\pm\pi$ (dashed) as a function of Zeeman term $V_Z$ (b). Gray regions mark the topological phase. Here $\Delta=0.1$ in (a).}
\label{PhaseDiagram.fig}
\end{figure}

In this appendix we provide additional analytical results supporting the main text conclusions for the spin-polarization of the 1D ferromagnetic system, and in particular in relation to the TPTs. 
A TPT is accompanied by gap closings at high symmetry points. Here we consider the BdG Hamiltonian of a 1D ferromagnetic system Eq.~\ref{BdG.eq}. It is straightforward to show that by increasing $V_Z$, the YSR band gap closes subsequently at $\Gamma$ ($k_x = 0$) and $M$ ($k_x = \pm \pi$) points in the first Brillouin zone \cite{Sato2006,Sato2010}. We plot the phase diagram for this Hamiltonian in Fig.~\ref{PhaseDiagram.fig}(a). 
By increasing $V_Z$ from zero and for $\mu<0$, first gap closing occurs at $k_x=0$ (green line) and the second gap closing occurs at $k_x=\pm\pi$ (yellow line), with the grey region in between being the topological phase. For $\mu>0$, only the order of the TPTs is inverted and therefore, without restricting our results, we assume $\mu<0$ in the following.
We stress here that along the lower critical coupling $V_Z^{c_-} = \sqrt{(|\mu|-2)^2+\Delta^2}$, the spin-polarization interchange of low-energy subgap states is in a one-to-one correspondence with the topological phase transition (TPT). 
The only exception is the point $\mu=0$, where the upper and lower critical couplings meet at $V_{Z} = \sqrt{4+\Delta^2}$, but there the topological phase shrinks to a point and thus does not exist. 
Therefore, we conclude that any spin-interchange of the bulk YSR states along the green line in Fig.~\ref{PhaseDiagram.fig}(a) is a clear identifier of TPT in 1D topological superconductors.

We continue our study by studying Eq.~\ref{BdG.eq} at high symmetry points. At the $\Gamma$ point the spin-orbit term, namely $L_R=2\lambda_R\sin{k_x}$, vanishes and the 1D Hamiltonian takes a particularly simple form
\be
\label{eq:H1Dsimple}
{\cal H}_{\rm 1D}(\Gamma) = \left( {\begin{array}{*{20}{c}}
{{\xi _0} + {V_Z}}&0&\Delta &0\\
0&{{\xi _0} - {V_Z}}&0&\Delta \\
\Delta &0&{ - {\xi _0} + {V_Z}}&0\\
0&\Delta &0&{ - {\xi _0} - {V_Z}}
\end{array}} \right),
\ee
where $\xi_0=-2-\mu$ is the kinetic energy at $k_x=0$.
Diagonalizing the Hamiltonian Eq.~\eqref{eq:H1Dsimple}, we find two spin-up and two spin-down eigenstates 
\be
\begin{array}{l}
\left| {E_ \uparrow ^ \pm } \right\rangle  = \left( {\begin{array}{*{20}{c}}
{ - \Delta }\\
0\\
{{\xi _0} \mp {\varepsilon_0 }}\\
0
\end{array}} \right)\frac{1}{{\sqrt {{\Delta ^2} + {{\left( {{\xi _0} \mp {\varepsilon_0}} \right)}^2}} }},\\
\left| {E_ \downarrow ^ \pm } \right\rangle   = \left( {\begin{array}{*{20}{c}}
0\\
\Delta \\
0\\
{{\xi _0} \mp {\varepsilon_0 }}
\end{array}} \right)\frac{1}{{\sqrt {{\Delta ^2} + {{\left( {{\xi _0} \mp {\varepsilon_0 }} \right)}^2}} }},
\end{array}
\ee
where we define ${\varepsilon_0} \equiv \sqrt {\xi _0^2 + {\Delta ^2}} $ and the eigenvalues are given by $E^{\pm}_{\uparrow} = V_Z \pm {\varepsilon_0}$ and $E^{\pm}_{\downarrow}= -V_Z \pm {\varepsilon_0}$. We depict the eigenvalues in Fig.~\ref{PhaseDiagram.fig}(b). 
Focusing on the lowest energy states, $E_{\downarrow}^{+}$ and $E_{\uparrow}^{-}$ cross each other at zero energy at $V_Z = \varepsilon_0$. The absolute values of other two eigenvalues increases with increasing $V_Z$, and thus never enter the subgap region. In the same fashion, we find the eigenstates of the Hamiltonian Eq.~\ref{BdG.eq} at $k_x=\pm\pi$ where again two branches enter the subgap region, shown in dashed lines in Fig.~\ref{PhaseDiagram.fig}(b). 

\begin{figure}[b]
\center
\includegraphics[width=0.4\textwidth]{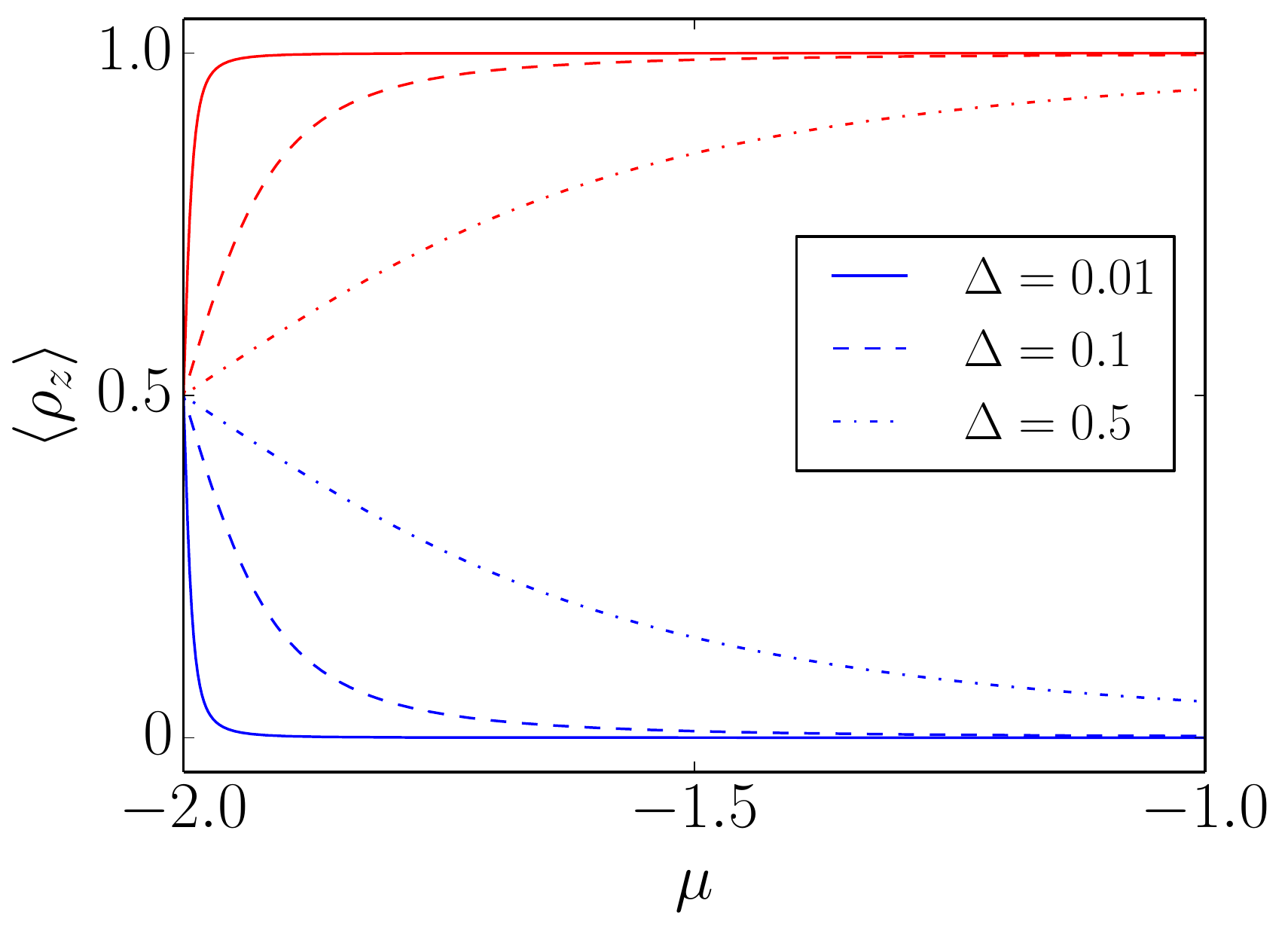}
\caption{Spin-polarization of YSR bands at the $\Gamma$ point for $\ket{E_{\downarrow}^{+}}$ (blue) and $\ket{E_{\uparrow}^{-}}$ (red).}
\label{etha.fig}
\end{figure}

We next concentrate on the electronic part of the wave function and evaluate the expectation value of the spin operator along the $\hat{z}$-direction given by $\rho_z = \sigma_z (\tau_0+\tau_z)/2$ for the two states we are interested in:
\be
\begin{array}{l}
\left\langle {E_\uparrow^-}\right| \rho_z {\left| {E_\uparrow^- }\right\rangle}
  = \frac{{{\Delta ^2}}}{{{\Delta ^2} + {{\left( {{\xi _0} + \sqrt {\xi _0^2 + {\Delta ^2}} } \right)}^2}}};\\
\left\langle {E_\downarrow^+}\right| \rho_z {\left| {E_\downarrow^+ }\right\rangle}
 = \frac{{ - {\Delta ^2}}}{{{\Delta ^2} + {{\left( {{\xi _0} - \sqrt {\xi _0^2 + {\Delta ^2}} } \right)}^2}}}.
\label{rho_z.eq}
\end{array}
\ee

Although the energy of these two eigenstates varies with $V_Z$, their spin-polarization does not depend on $V_Z$. We plot the spin-polarization given by Eq.~\ref{rho_z.eq} in Fig.~\ref{etha.fig} as a function of the chemical potential $\mu$ and for several different values of $\Delta$. The figure clearly shows that for a chemical potential at the bottom of the normal band, i.e.~$\mu=-2$, both subgap states acquire a finite electronic spin-polarization. However, moving away from the bottom of the band, the spin-up state (red) becomes fully electron-like and consequently fully spin-polarized, while the spin-down state (blue) becomes fully hole-like and thus rapidly loses its spin-polarization. In fact, for smaller, and thus more realistic, $\Delta$ this change in spin-polarization is even sharper.  

As a consequence of the spin-polarization being only dependent on the chemical potential $\mu$ and superconducting order parameter $\Delta$, we define a new variable $\eta = |2+\mu|/\Delta$ for which we identify two limits:
\begin{enumerate}
\item $\eta \ll 1$: Both positive- and negative-energy YSR bands have finite spin-polarization, with opposite spin-orientations. Fig.~2 in the main text belongs to this case, since there the chemical potential is $\mu = -2$, and thus $\eta = 0$.

\item $\eta \gg 1$: Only the spin-up state is dominantly electron-like and acquires a large spin-polarization, while the other state is dominantly hole-like, thus achieving only very minor spin-polarization. Fig.~3 in the main text belongs to this case, since $\mu = -1.85$ and $\eta = 15$.
\end{enumerate}
Since the denominator of $\eta$ is $\Delta$, which is generally the by far smallest energy scale in the problem, these are the only two relevant limits and values in-between would generally require extreme fine-tuning of the chemical potential.

\section{Ferromagnetic impurity chain}
\label{App2}
\begin{figure}[thb]
\includegraphics[width=0.5\textwidth]{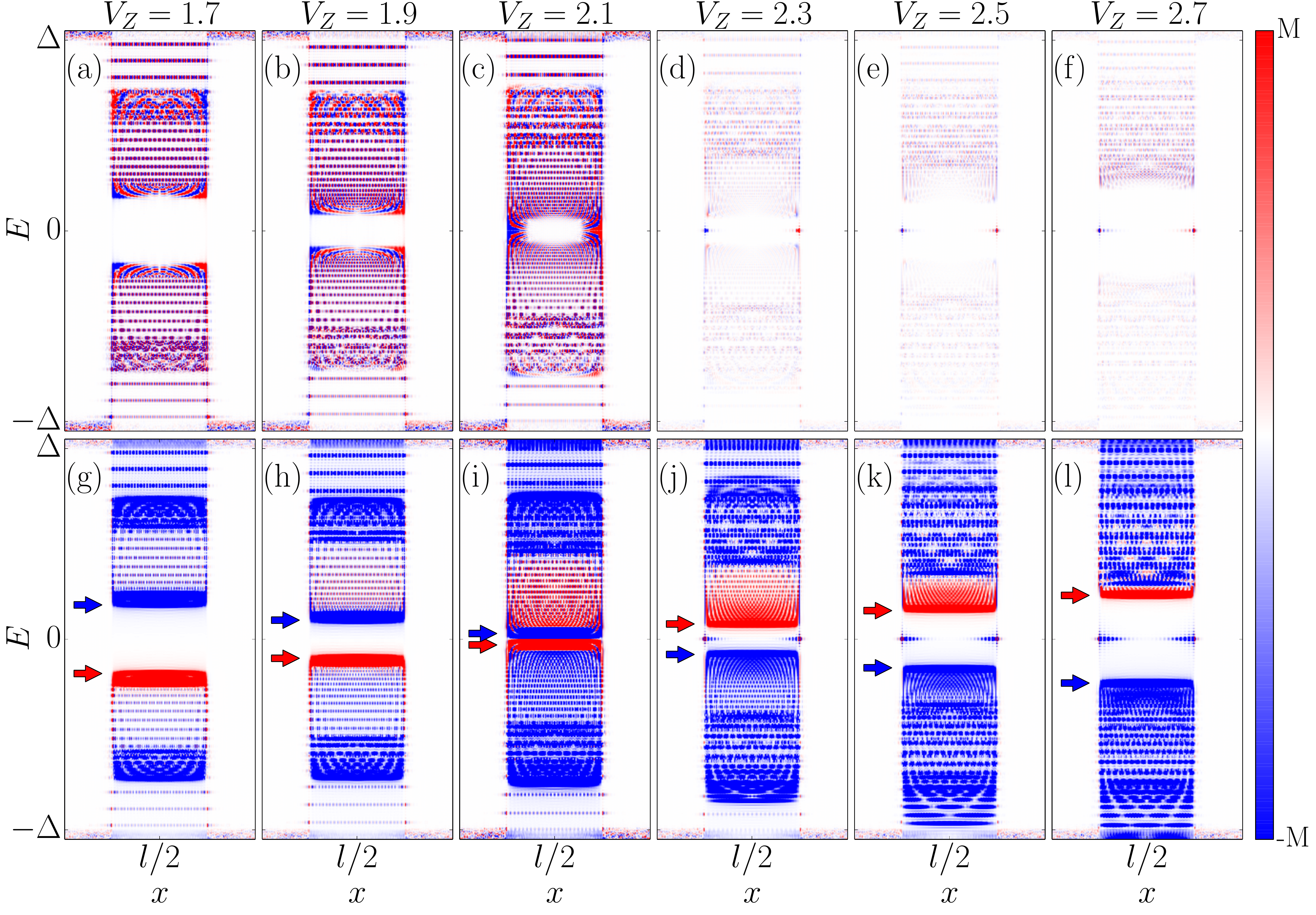}
\caption{SP-LDOS along a FMC for a gradual increase of $V_Z$ across the first TPT, with spin-polarization along ${x}$-axis (a-f) and $z$-axis (g-l). Here $\Delta_s = 0.4$, $\mu = -2.0$. Colored arrows mark the relevant spin-polarization of the low energy states. Color of the lower arrow always signals the topological phase in relation to the first TPT.}
\label{S1.fig}
\end{figure}

\begin{figure}[htb]
\includegraphics[width=0.5\textwidth]{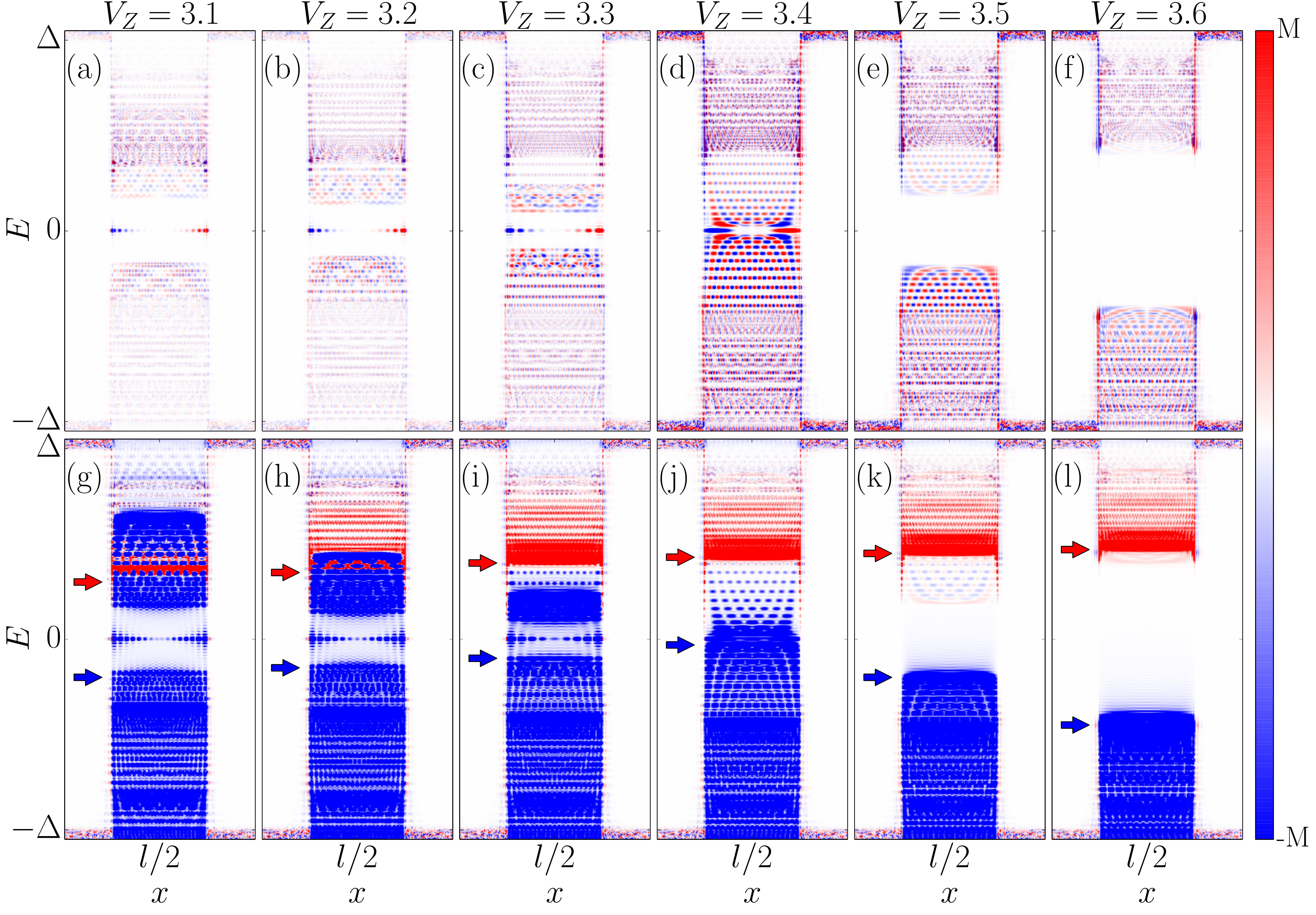}
\caption{Similar to Fig.~\ref{S1.fig}, except for larger $V_Z$ across the second TPT.}
\label{S2.fig}
\end{figure}

In this appendix we provide additional data on the spin-polarization for a dense FMC. In particular, we show in Fig.~\ref{S1.fig} and Fig.~\ref{S2.fig} the spin-polarized local density of states (SP-LDOS) for a gradual increase of the magnetic impurity moment $V_Z$, obtained from the lattice calculations. In Fig.~\ref{S1.fig} we trace through the transition from the topologically trivial into the non-trivial phase at $V_Z = 2.1$ and we plot both the SP-LDOS along the $x$-axis (\ref{S1.fig}(a-f)) and $z$-axis (\ref{S1.fig}(g-l)). 

In the trivial phase, the negative low-energy subgap states possess a spin-up polarization along the $z$-axis and by increasing $V_Z$ these states approach the Fermi level, see Figs.~\ref{S1.fig}(g-i). At the TPT, these states finally cross the Fermi level and go to positive energy, see Figs.~\ref{S1.fig}(j-l). The low-energy spin-down states have a complete reversed behavior, where they start from positive energy in the trivial phase and move on to negative energy in the topological phase. Therefore, the spin-polarization of both negative and positive low-energy states shows a spin-interchange across the TPT. This effect coincides with the appearance of Majorana bound states (MBSs) in the topologically non-trivial phase at the end-points of the impurity chain (see Figs.~\ref{S1.fig}(d-f)).

If we continue increasing the Zeeman field $V_Z$ further, as plotted in Fig.~\ref{S2.fig}, the spin-up polarized states move up to higher energies while some spin-down states with positive energy move down towards the Fermi level, see Figs.~\ref{S2.fig}(g-i). Therefore, in the topological phase, the positive-energy states close to Fermi level are not necessarily spin-up polarized. However, the negative-energy states remain spin-down polarized, and thus still clearly signal the topological phase. Finally, at $V_Z \sim 3.4$, the second TPT from topological into trivial phase occurs, see Fig.~\ref{S2.fig}(j). After the second TPT, there are no MBSs at the end-points of the impurity chain and also almost all the YSR states at positive (negative) energies are spin-up (spin-down) polarized. Thus for the second TPT the spin-polarization cannot be used to determine the topological phase. This is however not a limitation, since this regime requires such large magnetic moments as to completely suppress superconductivity.

\bibliography{bibFile.bib}

\begin{thebibliography}{56}
\bibitem [{Two()}]{Two}%
  \BibitemOpen
  \href@noop {} {\bibinfo  {journal} {Without loss of generality, we assume
  $\mu<0$ as for $\mu>0$ we only need to interchange spin-up and spin-down}\
  }\BibitemShut {NoStop}%
\bibitem [{Thr()}]{Three}%
  \BibitemOpen
  \href@noop {} {\bibinfo  {journal} {The closing of $\Delta_2$ happens only at
  a second TPT at $k_x=\pm\pi$ and for $V_Z=\sqrt{\Delta^2+(\mu+2)^2} \approx
  2$, which is in fact so large that it destroys superconductivity}\
  }\BibitemShut {NoStop}%
\bibitem [{One()}]{One}%
  \BibitemOpen
  \href@noop {} {\bibinfo  {journal} {Although the spin-helix is sufficient to
  reach the topological phase, we also include SOC in the SC substrate. We have
  confirmed that our SHC results are not sensitive to the SOC}\ }\BibitemShut
  {NoStop}%
\end{thebibliography}
\end{document}